\documentclass[showpacs,preprintnumbers,amsmath,amssymb,preprint]{revtex4}
\usepackage{graphicx}
\usepackage{dcolumn}
\usepackage{bm}

\def\ra{\rightarrow}
\newcommand{\GeV}{\ensuremath{\mathrm{Ge\kern -0.1em V}}}
\newcommand{\MeV}{\ensuremath{\mathrm{Me\kern -0.1em V}}}

\newcommand{\um}{\ensuremath{\mathrm{\mu m}}}
\newcommand{\pp}{\ensuremath{p\overline{p}}}
\newcommand{\pb}{\ensuremath{\mathrm{pb}^{-1}}}
\newcommand{\bs}{\ensuremath{B_s^0}}
\newcommand{\bd}{\ensuremath{B_d^0}}
\newcommand{\bsd}{\ensuremath{B_{s(d)}^0}}
\newcommand{\bu}{\ensuremath{B^{+}}}
\newcommand{\mm}{\ensuremath{\mu^{+}\mu^{-}}}

\newcommand{\hh}{\ensuremath{h^{+}h^{-}}}
\newcommand{\bsmm}{\ensuremath{\bs\ra\mm}}
\newcommand{\bdmm}{\ensuremath{\bd\ra\mm}}
\newcommand{\bsdmm}{\ensuremath{\bsd\ra\mm}}
\newcommand{\bjk}{\ensuremath{\bu\ra J/\psi K^{+}}}
\newcommand{\brbsmm}{\ensuremath{\mathcal{B}(\bsmm)}}
\newcommand{\brbdmm}{\ensuremath{\mathcal{B}(\bdmm)}}
\newcommand{\Mmm}{\ensuremath{M_{\mm}}}
\newcommand{\Lxy}{\ensuremath{\vec{L}_{T}}}
\newcommand{\ctau}{\ensuremath{\lambda}}
\newcommand{\pting}{\ensuremath{\Delta\Phi}}
\newcommand{\iso}{\ensuremath{\mathit{I}}}
\newcommand{\ptmm}{\ensuremath{\vec{p}^{\:\mm}_{T}}}
\newcommand{\cdf}{CDF~II}
\newcommand{\svx}{SVX~II}

\begin{document}
% ============================================================================
% --- TITLE PAGE
% ============================================================================

%%%\begin{flushright}
%%%CDF/PUB/BOTTOM/6875 \\
%%%submitted to PRL
%%%\end{flushright}
%%%\vspace*{0.50in}

\begin{center}
{\bf{ Search for \bsmm\ and \bdmm\ Decays \\
        in $p\overline{p}$ Collisions at $\sqrt{s}=1.96$~TeV }}
\end{center}

\font\eightit=cmti8
\def\r#1{\ignorespaces $^{#1}$}
\hfilneg
\begin{sloppypar}
\noindent
D.~Acosta,\r {15} T.~Affolder,\r 8 T.~Akimoto,\r {53}
M.G.~Albrow,\r {14} D.~Ambrose,\r {42} S.~Amerio,\r {41}  
D.~Amidei,\r {32} A.~Anastassov,\r {49} K.~Anikeev,\r {30} A.~Annovi,\r {43}
% ...
\end{sloppypar}
\vskip .026in
\begin{center}
(CDF Collaboration)
\end{center}

\vskip .026in
\begin{center}
\r 1  {\eightit Institute of Physics, Academia Sinica, Taipei, Taiwan 11529, 
Republic of China} \\
\r 2  {\eightit Argonne National Laboratory, Argonne, Illinois 60439} \\
\r 3  {\eightit Istituto Nazionale di Fisica Nucleare, University of Bologna,
I-40127 Bologna, Italy} \\
\r 4  {\eightit Brandeis University, Waltham, Massachusetts 02254} \\
\r 5  {\eightit University of California at Davis, Davis, California  95616} \\
\r 6  {\eightit University of California at Los Angeles, Los 
Angeles, California  90024} \\
\r 7  {\eightit University of California at San Diego, La Jolla, California  92093} \\ 
\r 8  {\eightit University of California at Santa Barbara, Santa Barbara, California 
93106} \\ 
\r 9 {\eightit Instituto de Fisica de Cantabria, CSIC-University of Cantabria, 
39005 Santander, Spain} \\
\r {10} {\eightit Carnegie Mellon University, Pittsburgh, PA  15213} \\
\r {11} {\eightit Enrico Fermi Institute, University of Chicago, Chicago, 
Illinois 60637} \\
\r {12}  {\eightit Joint Institute for Nuclear Research, RU-141980 Dubna, Russia}
\\
\r {13} {\eightit Duke University, Durham, North Carolina  27708} \\
\r {14} {\eightit Fermi National Accelerator Laboratory, Batavia, Illinois 
60510} \\
\r {15} {\eightit University of Florida, Gainesville, Florida  32611} \\
\r {16} {\eightit Laboratori Nazionali di Frascati, Istituto Nazionale di Fisica
               Nucleare, I-00044 Frascati, Italy} \\
\r {17} {\eightit University of Geneva, CH-1211 Geneva 4, Switzerland} \\
\r {18} {\eightit Glasgow University, Glasgow G12 8QQ, United Kingdom}\\
\r {19} {\eightit Harvard University, Cambridge, Massachusetts 02138} \\
\r {20} {\eightit The Helsinki Group: Helsinki Institute of Physics; and Division of
High Energy Physics, Department of Physical Sciences, University of Helsinki, FIN-00044, Helsinki, Finland}\\
\r {21} {\eightit Hiroshima University, Higashi-Hiroshima 724, Japan} \\
\r {22} {\eightit University of Illinois, Urbana, Illinois 61801} \\
\r {23} {\eightit The Johns Hopkins University, Baltimore, Maryland 21218} \\
\r {24} {\eightit Institut f\"{u}r Experimentelle Kernphysik, 
Universit\"{a}t Karlsruhe, 76128 Karlsruhe, Germany} \\
\r {25} {\eightit High Energy Accelerator Research Organization (KEK), Tsukuba, 
Ibaraki 305, Japan} \\
\r {26} {\eightit Center for High Energy Physics: Kyungpook National
University, Taegu 702-701; Seoul National University, Seoul 151-742; and
SungKyunKwan University, Suwon 440-746; Korea} \\
\r {27} {\eightit Ernest Orlando Lawrence Berkeley National Laboratory, 
Berkeley, California 94720} \\
\r {28} {\eightit University of Liverpool, Liverpool L69 7ZE, United Kingdom} \\
\r {29} {\eightit University College London, London WC1E 6BT, United Kingdom} \\
\r {30} {\eightit Massachusetts Institute of Technology, Cambridge,
Massachusetts  02139} \\   
\r {31} {\eightit Institute of Particle Physics, McGill University,
Montr\'{e}al, Canada H3A~2T8; and University of Toronto, Toronto, Canada
M5S~1A7} \\
\r {32} {\eightit University of Michigan, Ann Arbor, Michigan 48109} \\
\r {33} {\eightit Michigan State University, East Lansing, Michigan  48824} \\
\r {34} {\eightit Institution for Theoretical and Experimental Physics, ITEP,
Moscow 117259, Russia} \\
\r {35} {\eightit University of New Mexico, Albuquerque, New Mexico 87131} \\
\r {36} {\eightit Northwestern University, Evanston, Illinois  60208} \\
\r {37} {\eightit The Ohio State University, Columbus, Ohio  43210} \\  
\r {38} {\eightit Okayama University, Okayama 700-8530, Japan}\\  
\r {39} {\eightit Osaka City University, Osaka 588, Japan} \\
\r {40} {\eightit University of Oxford, Oxford OX1 3RH, United Kingdom} \\
\r {41} {\eightit University of Padova, Istituto Nazionale di Fisica 
          Nucleare, Sezione di Padova-Trento, I-35131 Padova, Italy} \\
\r {42} {\eightit University of Pennsylvania, Philadelphia, 
        Pennsylvania 19104} \\   
\r {43} {\eightit Istituto Nazionale di Fisica Nucleare, University and Scuola
               Normale Superiore of Pisa, I-56100 Pisa, Italy} \\
\r {44} {\eightit University of Pittsburgh, Pittsburgh, Pennsylvania 15260} \\
\r {45} {\eightit Purdue University, West Lafayette, Indiana 47907} \\
\r {46} {\eightit University of Rochester, Rochester, New York 14627} \\
\r {47} {\eightit The Rockefeller University, New York, New York 10021} \\
\r {48} {\eightit Istituto Nazionale di Fisica Nucleare, Sezione di Roma 1,
University di Roma ``La Sapienza," I-00185 Roma, Italy}\\
\r {49} {\eightit Rutgers University, Piscataway, New Jersey 08855} \\
\r {50} {\eightit Texas A\&M University, College Station, Texas 77843} \\
\r {51} {\eightit Texas Tech University, Lubbock, Texas 79409} \\
\r {52} {\eightit Istituto Nazionale di Fisica Nucleare, University of Trieste/\
Udine, Italy} \\
\r {53} {\eightit University of Tsukuba, Tsukuba, Ibaraki 305, Japan} \\
\r {54} {\eightit Tufts University, Medford, Massachusetts 02155} \\
\r {55} {\eightit Waseda University, Tokyo 169, Japan} \\
\r {56} {\eightit Wayne State University, Detroit, Michigan  48201} \\
\r {57} {\eightit University of Wisconsin, Madison, Wisconsin 53706} \\
\r {58} {\eightit Yale University, New Haven, Connecticut 06520} \\
\end{center}
 
%%%\author{CDF Authors}
%%%\affiliation{CDF Institutions}

\begin{center}
 (Received 19 March 2004) \\
 (published Phys.\ Rev.\ Lett.\ {\bf{93}}, 032001 (2004))
\end{center}

\begin{abstract}
  We report on a search for \bsmm\ and \bdmm\ decays in \pp\ collisions
  at $\sqrt{s} = 1.96$~TeV using $171\:\pb$ of data collected by the \cdf\ 
  experiment at the Fermilab Tevatron Collider. The decay rates of
  these rare processes are sensitive to contributions from physics beyond the 
  Standard Model.   One event survives all our selection requirements, 
  consistent with the background expectation.  We derive 
  branching ratio limits of $\brbsmm < 5.8\times10^{-7}$ and 
  $\brbdmm < 1.5\times10^{-7}$ at $90\%$ confidence level.  
\end{abstract}

\pacs{14.40.Nd, 13.20.He}

\maketitle

% ============================================================================
% --- BODY OF PAPER
% ============================================================================

The rare flavor-changing neutral current decay \bsmm ~\cite{qconj} is 
one of the most sensitive probes to physics beyond the Standard 
Model (SM)~[2-6].  The decay has not been observed and is currently limited 
to $\brbsmm < 2.0\times 10^{-6}$ at $90\%$ confidence level 
(CL)~\cite{bsmmruni}, while the SM prediction is 
$(3.5\pm 0.9)\times10^{-9}$~\cite{smbr}.  The limit on the related branching 
ratio, $\brbdmm < 1.6\times 10^{-7}$~\cite{belle},  is approximately 1000
times larger than its SM expectation.  The \brbsmm\ can be significantly 
enhanced in various supersymmetric (SUSY) extensions of the SM.  Minimal 
supergravity models at large $\tan\beta$~[3-5] predict 
$\brbsmm\leq\mathcal{O}(10^{-7})$ in regions of parameter space consistent 
with the observed muon $g-2$~\cite{gminus2} and also with the observed relic 
density of cold dark matter~\cite{wmap}.  $SO(10)$ models~\cite{so10}, which
naturally accommodate neutrino masses, predict a branching ratio as large as 
$10^{-6}$ in regions of parameter space consistent with these same 
experimental constraints.  $R$-parity violating SUSY models can also 
accommodate \brbsmm\ up to $10^{-6}$~\cite{arnowitt}.  Correspondingly, the 
\brbdmm\ can be enhanced by the same models.  Even modest improvements to 
the experimental limits can significantly restrict the available parameter 
space of these models.

We report on a search for \bsmm\ and \bdmm\ decays using the upgraded 
Collider Detector at Fermilab (\cdf) at the Tevatron \pp\ collider.   
The \cdf\ detector consists of a magnetic spectrometer surrounded by 
calorimeters and muon chambers and is described in detail in Ref.~\cite{TDR}.
A cylindrical drift chamber (COT) provides 96 measurement layers, 
organized into alternating axial and $\pm2^{\circ}$ stereo 
superlayers~\cite{COT}, and  a five-layer silicon microstrip detector 
(\svx) provides precise tracking information near the 
beamline~\cite{SVX}.   These are immersed in a $1.4$~T magnetic field and
measure charged particle momenta in the plane transverse to the beamline, 
$p_{T}$. Four layers of planar drift chambers (CMU) detect muons which 
penetrate the five absorption lengths of calorimeter steel~\cite{CMU}.  
Another four layers of planar drift chambers (CMP) instrument $0.6$~m of steel 
outside the magnet return yoke~\cite{CMP}.  The CMU and CMP chambers 
each provide coverage in the pseudo-rapidity range $\left|\eta\right|<0.6$,
where $\eta = -\ln(\tan\frac{\theta}{2})$ and $\theta$ is the angle of the 
track with respect to the beamline.  
%Cherenkov counters~\cite{CLC},
%covering the pseudo-rapidity range $3.7<\left|\eta\right|<4.7$, 
%measure the inelastic interaction rate used to estimate the integrated 
%luminosity, $\mathcal{L}$~\cite{LUMI}.
The dataset reported here corresponds to an integrated luminosity of 
$\mathcal{L}=171\pm10\;\pb$~\cite{LUMI}.

The data used in this analysis are selected by dimuon triggers. 
Muons are reconstructed as track stubs in the CMU chambers.  Two 
well-separated stubs are required and each is matched to a track reconstructed 
online using COT axial information~\cite{XFT}. The 
matched tracks must have $p_{T}>1.5\;\GeV/c$. A complete event reconstruction 
performed online confirms the $p_{T}$ and track-stub matching 
requirements.  If the overlapping CMP chambers contain a confirming muon
stub, the track is required to have $p_{T}>3\;\GeV/c$.  The 
two tracks must originate from the same vertex, be oppositely charged, and 
have an opening angle inconsistent with a cosmic ray event.  The invariant 
mass of the muon pair must satisfy $\Mmm < 6\;\GeV/c^{2}$.  Events in which 
neither muon is reconstructed with a CMP stub must additionally satisfy 
$p_{T}^{\mu^+}+p_{T}^{\mu^-}>5\;\GeV/c$ and $\Mmm > 2.7\;\GeV/c^{2}$.  
This set of triggers is used for all the data included here and
events passing these requirements are recorded for further analysis.

Our offline analysis begins by identifying the muon candidates and matching
them to the trigger tracks using COT hit information.  To avoid regions of 
rapidly changing trigger efficiency, we omit muons with $p_{T}<2\;\GeV/c$.  
To reduce backgrounds from fake muons, stricter track-stub matching 
requirements are made and the vector sum of the muon momenta must satisfy 
$| \ptmm | > 6\;\GeV/c$.  To ensure good vertex 
resolution, stringent requirements are made on the number of \svx\ hits 
associated with each track.  Surviving events have the two muon tracks 
constrained to a common 3-D vertex satisfying vertex quality 
requirements.  The two-dimensional decay length, $| \Lxy |$, is calculated 
as the transverse distance from the beamline to the dimuon vertex and is
signed relative to \ptmm. For each $B$-candidate we estimate the proper decay 
length using 
$\ctau=c\, \Mmm | \Lxy | / | \ptmm |$.  In the data, 2981 events survive all 
the above trigger and offline reconstruction requirements.  This forms a 
background-dominated sample with contributions from two principal sources: 
combinatoric background events with a fake muon and events from generic 
$B$-hadron decays (e.g. sequential semi-leptonic 
decays $b\rightarrow c\mu^{-}X\rightarrow \mm X$ or double 
semi-leptonic decay in gluon splitting events 
$g\rightarrow b\overline{b}\rightarrow\mm X$).
%
%%%Using the best previously published limit as an estimate 
%%%for the branching ratio, we expect at most about 28 (9) \bsdmm\ decays to 
%%%survive these cuts.

We model the signal decays using the Pythia Monte Carlo (MC)~\cite{pythia} 
tuned to inclusive $B$-hadron data~\cite{lannon}.  The Pythia events are 
passed through a full detector simulation and satisfy the same requirements 
as data.  To normalize to experimentally determined cross-sections, we require 
$p_{T}(\bsd)>6\;\GeV/c$ and rapidity $\left| y \right| < 1$.

\begin{figure}
  \includegraphics[width=8.0cm]{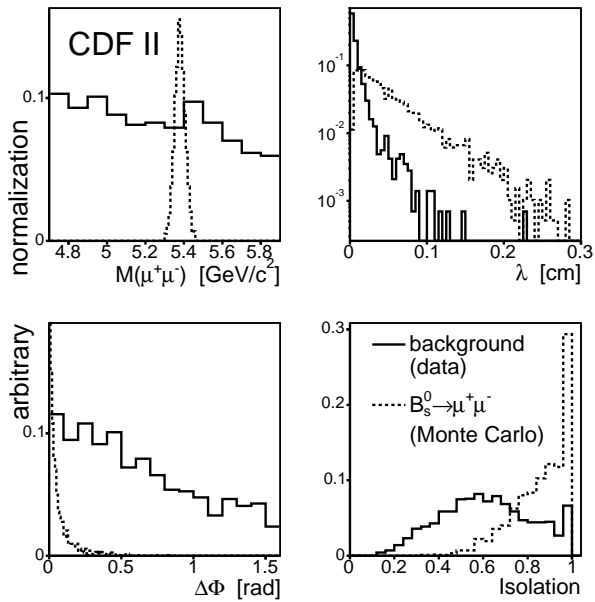}
  \caption{\label{fig:SvB} Arbitrarily normalized distributions of the 
    discriminating variables for events in our background-dominated data 
    sample (solid) compared to Monte Carlo \bsmm\ events (dashed).}
\end{figure}

To discriminate \bsdmm\ decays from background events we use these four 
variables: 
the invariant mass of the muon pair (\Mmm);
the $B$-candidate proper decay length (\ctau); 
the opening angle (\pting) between the $B$-hadron flight direction 
(estimated as the vector \ptmm) and the vector \Lxy; 
and the $B$-candidate track isolation (\iso)~\cite{iso}.
Figure~\ref{fig:SvB} shows the distributions of these variables for 
background-dominated data and MC signal events.  

A ``blind'' analysis technique is used to determine the optimal selection 
criteria for these four variables.  The
data in the search window $5.169 < \Mmm < 5.469\:\GeV/c^2$  
are hidden and the optimization performed using only data in the sideband 
regions, $4.669 < \Mmm < 5.169\:\GeV/c^2$ and $5.469 < \Mmm < 5.969\:\GeV/c^2$.
The search region corresponds to approximately $\pm4$ times the two-track 
invariant mass resolution centered on the \bs\ and \bd\ masses~\cite{masses}.
We use the set of (\Mmm, \ctau, \pting, \iso) criteria which minimizes the 
{\it{a priori}} expected $90\%$~CL upper limit on the branching ratio.  For a 
given number of observed events, $n$, and an expected background of 
$n_{\mathrm{bg}}$, the branching ratio is determined using:
%$
\begin{equation*}
\brbsmm \leq \frac{N(n,n_{\mathrm{bg}})}{2\,\sigma_{\bs}\,\mathcal{L}\,\alpha\,\epsilon_{\mathrm{total}}}
\end{equation*}
%$,
where $N(n,n_{\mathrm{bg}})$ is the number of candidate \bsmm\ decays at 90\% 
CL, estimated using the Bayesian approach of Ref.~\cite{bayes} 
and incorporating the uncertainties into the limit. 
The {\it {a priori}} expected limit is given by the sum 
over all possible observations, $n$, weighted by the corresponding Poisson 
probability when expecting $n_{\mathrm{bg}}$.  The \bs\ 
production cross-section is estimated as 
$\sigma_{\bs}=\frac{f_s}{f_u}\sigma_{B^+}$, where 
$\frac{f_s}{f_u}=\frac{0.100}{0.391}$~\cite{HFWG} and $\sigma_{B^+}$ is taken 
from Ref.~\cite{BXS}.  
For the \bdmm\ limit we substitute $\sigma_{\bd}$ for $\sigma_{\bs}$, 
$f_d$ for $f_s$, and assume $f_d = f_u$.
The factor of two in the denominator accounts for the charge-conjugate 
$B$-hadron final states.  The expected background, $n_{\mathrm{bg}}$, and
the total acceptance times efficiency, $\alpha\,\epsilon_{\mathrm{total}}$,
are estimated separately for each combination of requirements.  

For both signal and background, the variables \ctau\ and \pting\ are the only
correlated variables with a linear correlation coefficient 
of $-0.3$.  Thus we estimate the number of background events as 
$n_{\mathrm{bg}} = n_{\mathrm{sb}}(\ctau,\pting)\, f_{\iso}\, f_{M}$, 
where $n_{\mathrm{sb}}(\ctau,\pting)$ is the number of sideband events passing 
a particular set of \ctau\ and \pting\ cuts, $f_{\iso}$ is the fraction of 
background events that survive a given \iso\ requirement, and 
$f_{M}$ is the ratio of the number of events in the search window to the 
number of events in the sideband regions.  Since \Mmm\ and \iso\ are 
uncorrelated with the rest of the variables, we evaluate $f_{M}$ and 
$f_{\iso}$ on samples with no \ctau\ or \pting\ 
requirement, thus reducing their associated uncertainty. 

We estimate $f_{\iso}$ from the background-dominated sample for a variety
of thresholds.   We investigate sources of systematic bias by
calculating $f_{\iso}$ in bins of \Mmm\ and \ctau\ and conservatively 
assign a relative systematic uncertainty of $\pm 5\%$.
Since the \Mmm\ distribution of the background-dominated sample is well 
described by a first-order polynomial, $f_{M}$ is given by the 
ratio of widths of the search to sideband regions.

MC studies demonstrate that our estimate of $n_{\mathrm{bg}}$ 
accurately accounts for generic $b\overline{b}$ contributions, while two-body 
decays of $B$-mesons ($B^{0}_{s(d)}\ra\hh$, where $h^{\pm}=\pi^{\pm}$ or 
$K^{\pm}$) are estimated to contribute to the search region at levels at 
least 100 times smaller than our expected sensitivity.  

Using these background-dominated control samples, $\mu^{\pm}\mu^{\pm}$ 
events and \mm\ events with $\ctau<0$,   we compare our background predictions 
to the number of events observed in the search window for a wide range of 
(\ctau, \pting, \iso) requirements.   No statistically significant 
discrepancies are observed.  For example, using the optimized set 
of selection criteria described below and summing over these control samples, 
we get a total prediction of $3\pm1$ events and observe five.  Another 
cross-check is performed using a fake muon enhanced \mm\ sample. 
By requiring at least one of the muon legs to fail the muon identification 
requirements, we reduce the signal efficiency by a factor of 50 while 
increasing the background acceptance by a factor of three.  In this sample, 
using the optimized requirements, we predict $6\pm1$ and observe seven events.

We estimate the total acceptance times efficiency as 
$\alpha\,\epsilon_{\mathrm{total}}=\alpha\,\epsilon_{\mathrm{trig}}\,\epsilon_{\mathrm{reco}}\,\epsilon_{\mathrm{final}}$, 
where $\alpha$ is the geometric and kinematic acceptance of the trigger, 
$\epsilon_{\mathrm{trig}}$ is the trigger efficiency for events in the
acceptance, $\epsilon_{\mathrm{reco}}$ is the offline reconstruction efficiency
for events passing the trigger, and $\epsilon_{\mathrm{final}}$ is the 
efficiency for passing the final cuts on the discriminating variables for 
events satisfying the trigger and reconstruction requirements.  For the 
optimization, only $\epsilon_{\mathrm{final}}$ changes as we vary the 
requirements on \Mmm, \ctau, \pting, and \iso.

The acceptance is estimated as the fraction of \bsdmm\ MC events 
which fall within the geometric acceptance and satisfy the kinematic
requirements of at least one of the analysis triggers.
We find $\alpha=(6.6\pm0.5)\%$.  The uncertainty includes roughly equal 
contributions from systematic variations of the modeling of the $B$-hadron 
$p_{T}$ spectrum and longitudinal beam profile, and from the statistics of the
sample.  It also includes negligible contributions from variations of the 
beamline offsets and of the detector material description used in the 
simulation.

The trigger efficiency, including the effects of the offline-to-trigger 
track matching, is estimated from samples of $J/\psi\ra\mm$ 
decays selected with a trigger requiring only one identified muon.  The data 
are used to parameterize the trigger efficiency as a function of $p_{T}$ 
and $\eta$ for the unbiased muon.  The efficiency for 
\bsdmm\ decays is determined by the convolution of this parameterization with 
the $(p_{T}^{\mu^{+}},\eta^{\mu^{+}},p_{T}^{\mu^{-}},\eta^{\mu^{-}})$ spectra 
of signal MC events within the acceptance.  Including the 
online reconstruction requirements, the trigger efficiency 
is $\epsilon_{\mathrm{trig}} = (85\pm3)\%$.  The uncertainty is dominated by 
the systematic uncertainty accounting for kinematic differences between 
$J/\psi\ra\mm$ and \bsdmm\ decays and also includes contributions 
from variations in the functional form used in the parameterization, 
the effects of two-track correlations, and sample statistics.

The offline reconstruction efficiency is given by the product
$\epsilon_{\mathrm{reco}}=\epsilon_{\mathrm{COT}}\,\epsilon_{\mu}\,\epsilon_{\mathrm{SVX}}$,
where $\epsilon_{\mathrm{COT}}$ is the absolute reconstruction efficiency of 
the COT, 
$\epsilon_{\mu}$ is the muon reconstruction efficiency given a COT track, 
and $\epsilon_{\mathrm{SVX}}$ is the fraction of reconstructed muons which 
satisfy the \svx\ requirements.  Each term is a two-track 
efficiency.  A hybrid data-MC method is used to determine 
$\epsilon_{\mathrm{COT}}$. Occupancy effects are accounted for
by embedding COT hits from MC tracks in data events.  The MC 
simulation is tuned at the hit level to reproduce residuals, hit width and 
hit usage in the data. For embedded muons with $p_{T}>2\;\GeV/c$, we measure 
$\epsilon_{\mathrm{COT}}=99\%$.  Using the unbiased $J/\psi\ra\mm$
samples, we estimate the muon reconstruction efficiency, including the 
track-stub matching requirements, to be $96\%$.
A sample of $J/\psi\ra\mm$ events satisfying our COT and
muon reconstruction requirements is used to determine 
$\epsilon_{\mathrm{SVX}}=75\%$.  
The total reconstruction efficiency is given by the above product, 
$\epsilon_{\mathrm{reco}}=(71\pm3)\%$.
The uncertainty is dominated by the systematic uncertainty accounting for 
kinematic differences between $J/\psi\ra\mm$ and \bsdmm\ decays and also 
includes contributions from the variation of the COT simulation
parameters and sample statistics.

The efficiency $\epsilon_{\mathrm{final}}$ is determined from the \bsdmm\ MC 
sample and varies from $28-78\%$ over the range of (\Mmm, \ctau, \pting, \iso)
requirements considered in the optimization.  The MC 
modeling is checked by comparing the mass resolution and \ctau,
\pting, and \iso\ efficiency as a function of selection threshold for 
\bjk($J/\psi\ra\mm$) events.  The \bjk\ MC sample is produced in the 
same manner as the \bsmm\ sample.  The \bjk\ data sample is collected using 
dimuon triggers very similar to those used in the analysis, but with
a larger acceptance for \bjk\ decays.  We make the same requirements on the 
dimuon tracks and vertex as employed in the analysis.  
The MC efficiency is consistent with the sideband-subtracted data efficiency 
for a range of cut thresholds within $5\%$ (relative),
which is assigned as a systematic uncertainty on $\epsilon_{\mathrm{final}}$. 
In both the data and the MC samples the mean of the three-track invariant mass 
distribution is within $3\:\MeV/c^2$ of the world average $B^+$ mass.  The 
two-track invariant mass resolution is well described by the MC.  

The optimal set of selection criteria uses a $\pm 80\;\MeV/c^{2}$ search 
window around the \bs\ mass, $\ctau>200\;\um$, $\pting < 0.10$~rad and 
$\iso > 0.65$.  The mass resolution, estimated from the MC for the events 
surviving all requirements, is $27\;\MeV/c^{2}$ so that the \bd\ and \bs\ 
masses are resolved. We define a separate search window centered on the world 
average \bd\ mass and use the same set of selection criteria for the \bdmm\ 
search.  The total acceptance times efficiency is 
$\alpha\,\epsilon_{\mathrm{total}}=(2.0\pm0.2)\%$ for both decays.

\begin{figure}[h]
  \includegraphics[width=8.0cm]{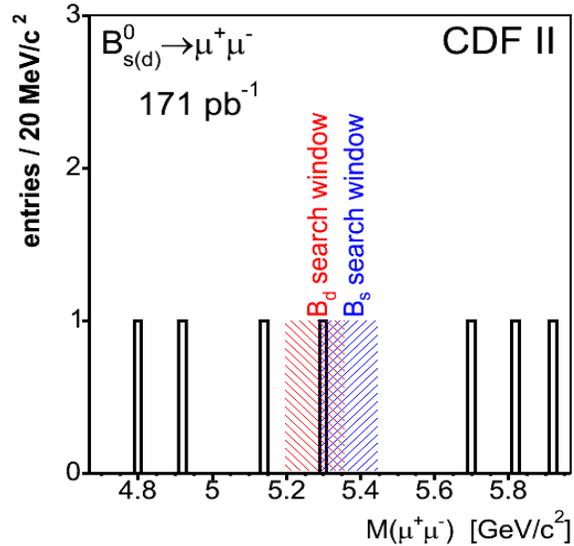}
  \caption{\label{fig:results} The \mm\ invariant mass distribution of the 
    events in the sideband and search regions satisfying all requirements.}
\end{figure}

Using these criteria one event survives 
all requirements and has an invariant mass of $\Mmm=5.295\;\GeV/c^{2}$, thus 
falling into both the \bs\ and \bd\ search windows as shown in 
Figure~\ref{fig:results}.  This is consistent with the $1.1\pm0.3$ background 
events expected in each of the \bs\ and \bd\ mass windows.
We derive 90\% (95\%) CL limits of
$\brbsmm < 5.8\times 10^{-7}$ $(7.5\times 10^{-7})$ and
$\brbdmm < 1.5\times 10^{-7}$ $(1.9\times 10^{-7})$.  
The new \bsmm\ limit improves the previous 
limit~\cite{bsmmruni} by a factor of three and significantly reduces the 
allowed parameter space of $R$-parity violating and $SO(10)$ SUSY
models~\cite{arnowitt,so10}.  The \bdmm\ limit is slightly better than 
the recent limit from the Belle Collaboration~\cite{belle}.
%%%We expect significant improvements to this analysis as we work to increase 
%%%the signal acceptance, reduce background contributions, and collect
%%%more data.

% ============================================================================
% --- ACKNOWLEDGEMENTS
% ============================================================================
We thank the Fermilab staff and the technical staffs of the participating 
institutions for their vital contributions. This work was supported by the 
U.S. Department of Energy and National Science Foundation; 
the Italian Istituto Nazionale di Fisica Nucleare; 
the Ministry of Education, Culture, Sports, Science and Technology of Japan; 
the Natural Sciences and Engineering Research Council of Canada; 
the National Science Council of the Republic of China; 
the Swiss National Science Foundation; 
the A.P. Sloan Foundation; 
the Bundesministerium fuer Bildung und Forschung, Germany; 
the Korean Science and Engineering Foundation and 
the Korean Research Foundation; 
the Particle Physics and Astronomy Research Council and the Royal Society, UK; 
the Russian Foundation for Basic Research; 
the Comision Interministerial de Ciencia y Tecnologia, Spain; and
in part by the European Community's Human Potential Programme under contract 
HPRN-CT-20002, Probe for New Physics.

% ============================================================================
% --- BIBLIOGRAPHY
% ============================================================================

%%%\bibliography{prl}

\begin{thebibliography}{99.}
\bibitem{qconj}
 Throughout this paper inclusion of charge conjugate modes is implicit.
\bibitem{iso}
 The $B$-candidate isolation is defined as 
 $\iso=| \ptmm | /(\sum_{i}p^{i}_{T} + | \ptmm | )$, where the
 sum is over all tracks within an $\eta$-$\phi$ cone radius of $1$, centered
 on \ptmm\ and satisfying standard track quality requirements; $\phi$ is the
 azimuthal angle defined in the plane transverse to the beamline.
\bibitem{masses} The mass windows are centered on $5.369\:\:\GeV/c^2$ 
  and $5.279\:\:\GeV/c^2$ for the \bs\ and \bd\ searches, respectively.
\end{thebibliography}

% ============================================================================
\end{document}